\documentclass[sigconf]{aamas}  
\usepackage{balance}  

\usepackage{booktabs}

\setcopyright{ifaamas}  
\acmDOI{}  
\acmISBN{}  
\acmConference[AAMAS'19]{Proc.\@ of the 18th International Conference on Autonomous Agents and Multiagent Systems (AAMAS 2019)}{May 13--17, 2019}{Montreal, Canada}{N.~Agmon, M.~E.~Taylor, E.~Elkind, M.~Veloso (eds.)}  
\acmYear{2019}  
\copyrightyear{2019}  
\acmPrice{}  

\usepackage{amsmath,amsfonts,bm}

\def\eqref#1{equation~\ref{#1}}

\def\1{\bm{1}}

\DeclareMathAlphabet{\mathsfit}{\encodingdefault}{\sfdefault}{m}{sl}
\SetMathAlphabet{\mathsfit}{bold}{\encodingdefault}{\sfdefault}{bx}{n}

\usepackage{subfig}
\usepackage{float}
\usepackage [english]{babel}
\usepackage [autostyle, english = american]{csquotes}
\MakeOuterQuote{"}
\usepackage{graphicx}
\usepackage{enumitem}
\setlist[itemize]{noitemsep}
\setlist[enumerate]{noitemsep}
\usepackage{hyperref}
\usepackage{xcolor}
\usepackage[normalem]{ulem}

\usepackage{amsmath}
\usepackage{algorithm}
\usepackage[noend]{algpseudocode}

\usepackage{bbm}

\usepackage{verbatim}

\usepackage{thmtools}
\usepackage{thm-restate}

\usepackage{ulem}

\begin{document}

\title{Evolving Intrinsic Motivations for Altruistic Behavior}



\author{Jane X. Wang, Edward Hughes, Chrisantha Fernando, Wojciech M. Czarnecki, \\
	Edgar A. Du\'e\~nez-Guzm\'an, \& Joel Z. Leibo}
\affiliation{%
 \institution{DeepMind}
 \city{London} 
 \postcode{N1C 4AG}
}
\email{{wangjane, edwardhughes, chrisantha, lejlot, duenez, jzl} @google.com}

\renewcommand{\shortauthors}{J. Wang et al.}

\begin{abstract}
Multi-agent cooperation is an important feature of the natural world. Many tasks involve individual incentives that are misaligned with the common good, yet a wide range of organisms from bacteria to insects and humans are able to overcome their differences and collaborate. Therefore, the emergence of cooperative behavior amongst self-interested individuals is an important question for the fields of multi-agent reinforcement learning (MARL) and evolutionary theory. Here, we study a particular class of multi-agent problems called intertemporal social dilemmas (ISDs), where the conflict between the individual and the group is particularly sharp. By combining MARL with appropriately structured natural selection, we demonstrate that individual inductive biases for cooperation can be learned in a model-free way. To achieve this, we introduce an innovative modular architecture for deep reinforcement learning agents which supports multi-level selection. We present results in two challenging environments, and interpret these in the context of cultural and ecological evolution.

\end{abstract}


\maketitle

\section{Introduction}

Nature shows a substantial amount of cooperation at all scales, from microscopic interactions of genomes and bacteria to species-wide societies of insects and humans \citep{smith1997major}. This is in spite of natural selection pushing for short-term individual selfish interests \citep{darwin1859}. In its purest form, altruism can be favored by selection when cooperating individuals preferentially interact with other cooperators, thus realising the rewards of cooperation without being exploited by defectors \citep{hamilton-1964a, hamilton1964genetical, dawkins1976selfish, santos2006cooperation, fletcher2009simple}.  However, many other possibilities exist, including kin selection, reciprocity and group selection \citep{Nowak1560, ubeda2011power, trivers1971evolution, nowak2005evolution, wilson1975theory, smith1964group}.  

Lately the emergence of cooperation among self-interested agents has become an important topic in multi-agent deep reinforcement learning (MARL). \cite{leibo17} and \cite{hughes2018inequity} formalize the problem domain as an {\em intertemporal social dilemma} (ISD), which generalizes matrix game social dilemmas to Markov settings. Social dilemmas are characterized by a trade-off between collective welfare and individual utility. As predicted by evolutionary theory, self-interested reinforcement-learning agents are typically unable to achieve the collectively optimal outcome, converging instead to defecting strategies \citep{leibo17, perolat17}.
The goal is to find multi-agent training regimes in which individuals resolve social dilemmas, i.e., cooperation emerges.
Previous work has found several solutions, belonging to three broad categories: 1) opponent modelling \citep{foerster17, KleimanWeiner2016}, 2) long-term planning using perfect knowledge of the game's rules  \citep{lerer17, peysakhovich2018} and 3) a specific intrinsic motivation function drawn from behavioral economics \citep{hughes2018inequity}. These hand-crafted approaches run at odds with more recent end-to-end model-free learning algorithms, which have been shown to have a greater ability to generalize (e.g. \citep{espeholt2018impala}). We propose that evolution can be applied to remove the hand-crafting of intrinsic motivation, similar to other applications of evolution in deep learning.

Evolution has been used to optimize single-agent hyperparameters \citep{jaderberg2017population}, implement black-box optimization \citep{wierstra2008natural}, and to evolve neuroarchitectures \citep{miller1989designing, stanley2002evolving}, regularization \citep{chan2002alleviating}, loss functions \citep{jaderberg2018human, houthooft2018evolved}, behavioral diversity \citep{conti2017improving}, and entire reward functions \citep{singh2009rewards, singh2010intrinsically}. These principles tend to be driven by single-agent search and optimization or competitive multi-agent tasks. Therefore there is no guarantee of success when applying them in the ISD setting. More closely related to our domain are evolutionary simulations of predator-prey dynamics \citep{yong2001cooperative}, which used enforced subpopulations to evolve populations of neurons which are sampled to form the hidden layer of a neural network.\footnote{See also \cite{potter2000cooperative} and \cite{panait2005cooperative} for reviews of other evolutionary approaches to cooperative multi-agent problems.}

To address the specific challenges of ISDs, the system we propose distinguishes between optimization processes that unfold over two distinct time-scales: (1) the fast time-scale of learning and (2) the slow time-scale of evolution \citep[similar to][]{hinton1987learning}. In the former, individual agents repeatedly participate in an intertemporal social dilemma using a fixed intrinsic motivation. In the latter, that motivation is itself subject to natural selection in a population. We model this intrinsic motivation as an additional additive term in the reward of each agent \citep{chentanez2005}. We implement the intrinsic reward function as a two-layer fully-connected feed-forward neural network, whose weights define the genotype for evolution. We propose that evolution can help mitigate this intertemporal dilemma by bridging between these two timescales via an intrinsic reward function.

Evolutionary theory predicts that evolving individual intrinsic reward weights across a population who interact uniformly at random does not lead to altruistic behavior \citep{axelrod81}. Thus, to achieve our goal, we must structure the evolutionary dynamics \citep{Nowak1560}. We first implement a ``Greenbeard'' strategy \citep{dawkins1976selfish, jansen2006} in which agents choose interaction partners based on an honest, real-time signal of cooperativeness. We term this process {\em assortative matchmaking}. Although there is ecological evidence of assortative matchmaking \citep{keller98}, it cannot explain cooperation in all taxa \citep{grafen1990animals, henrich2003, gardner2010greenbeards}. Moreover it isn't a general method for multi-agent reinforcement learning, since honest signals of cooperativeness are not normally observable in the ISD models typically studied in deep reinforcement learning.

\begin{figure*}[htb]
  \centering     
  \includegraphics[width=0.7\textwidth]{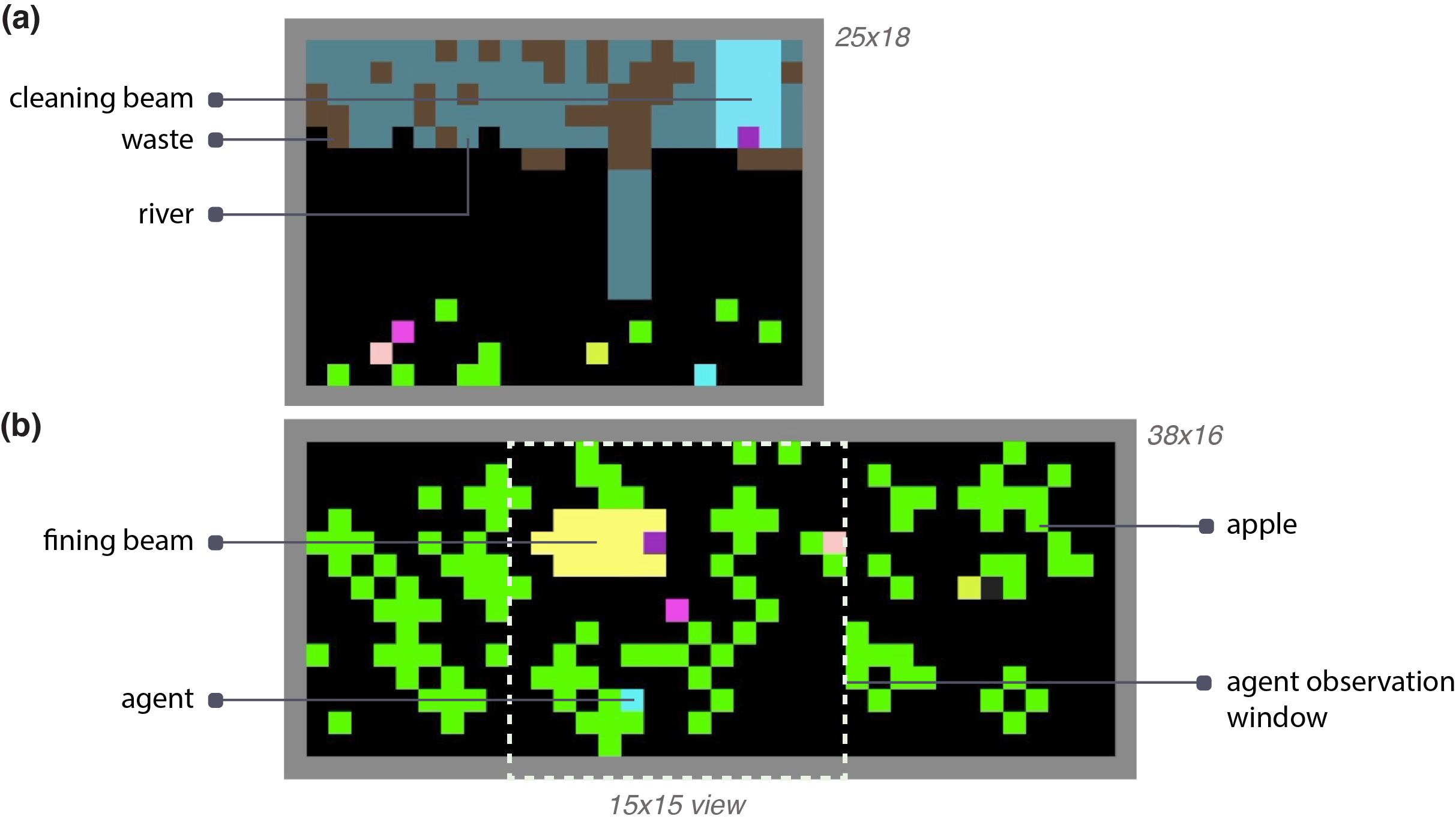}
  \vspace{-.25cm}
     \caption{Screenshots from (a) the Cleanup game, (b) the Harvest game. The size of the agent-centered observation window is shown in (b). The same size observation was used in all experiments.}
     \vspace{-.25cm}
     \label{fig:gallery}
\end{figure*}

To address the limitations of the assortative matchmaking approach, we introduce an alternative modular training scheme loosely inspired by ideas from the theory of multi-level (group) selection \citep{wilson1975theory, henrich2003}, which we term {\em shared reward network} evolution. Here, agents are composed of two neural network modules: a policy network and a reward network. On the fast timescale of reinforcement learning, the policy network is trained using the modified rewards specified by the reward network. On the slow timescale of evolution, the policy network and reward network modules evolve separately from one another. In each episode every agent has a distinct policy network but the same reward network. As before, the fitness for the policy network is the individual's reward. In contrast, the fitness for the reward network is the collective return for the entire group of co-players. In terms of multi-level selection theory, the policy networks are the lower level units of evolution and the reward networks are the higher level units. Evolving the two modules separately in this manner prevents  evolved reward networks from overfitting to specific policies. This evolutionary paradigm not only resolves difficult ISDs without handcrafting but also points to a potential mechanism for the evolutionary origin of social inductive biases.

The paper is structured as follows. 
In Section 2, we define our problem domain, and describe in detail our agent architecture and training methods. In Section 3, we present results from our experiments and further analyses of agent policies. Finally in Section 4, we discuss interpretations of our model as well as make suggestions for future work.

\section{Methods}

We varied and explored different combinations of parameters, namely: (1) environments \{Harvest, Cleanup\}, (2) reward network features \{prospective, retrospective\}, (3) matchmaking \{random, assortative\}, and (4) reward network evolution \{individual, shared, none\}. We describe these in the following sections.

\subsection{Intertemporal social dilemmas}

In this paper, we consider Markov games \citep{littman1994markov} within a MARL setting. Specifically we study intertemporal social dilemmas \citep{leibo17, hughes2018inequity}, defined as games in which individually selfish actions produce individual benefit on short timescales but have negative impacts on the group over a longer time horizon. This conflict between the two timescales characterizes the intertemporal nature of these games. The tension between individual and group-level rationality identifies them as social dilemmas (e.g. the famous Prisoner's Dilemma).  

We consider two dilemmas, each implemented as a partially observable Markov game on a 2D grid (see Figure \ref{fig:gallery}), with $N=5$ players playing at a time. In the \emph{Cleanup} game, agents tried to collect apples (reward ${+}1$) that spawned in a field at a rate inversely related to the cleanliness of a geographically separate aquifer. Over time, this aquifer filled up with waste, lowering the respawn rate of apples linearly, until a critical point past which no apples could spawn. Episodes were initialized with no apples present and zero spawning, thus necessitating cleaning. The dilemma occurred because in order for apples to spawn, agents must leave the apple field and clean, which conferred no reward. However if all agents declined to clean (defect), then no rewards would be received by any. In the \emph{Harvest} game, again agents collected rewarding apples. The apple spawn rate at a particular point on the map depended on the number of nearby apples, falling to zero once there were no apples in a certain radius. There is a dilemma between the short-term individual temptation to harvest all the apples quickly and the consequential rapid depletion of apples, leading to a lower total yield for the group in the long-term. 

All episodes last 1000 steps, and the total size of the playable area is 25$\times$18 for Cleanup and 38$\times$16 for Harvest. Games are partially observable in that agents can only observe via a 15$\times$15 RGB window, centered on their current location. The action space consists of moving left, right, up, and down, rotating left and right, and the ability to tag each other. This action has a reward cost of 1 to use, and causes the player tagged to lose 50 reward points, thus allowing for the possibility of punishing free-riders \citep{oliver1980rewards, gurerk2006competitive}. The Cleanup game has an additional action for cleaning waste.

\subsection{Modeling social preferences as intrinsic motivations}

In our model, there are three components to the reward that enter into agents' loss functions (1) total reward, which is used for the policy loss, (2) extrinsic reward, which is used for the extrinsic value function loss and (3) intrinsic reward, which is used for the intrinsic value function loss.

The {\em total reward} for player $i$ is the sum of the extrinsic reward and an intrinsic reward as follows:
\begin{align}\label{eq:totalreward}
    r_i(s_i,a_i) = r_i^E(s_i,a_i) + u_i(\mathbf{f}_i) \, .
\end{align}
The {\em extrinsic reward} $r^E_i(s, a)$ is the environment reward obtained by player $i$ when it takes action $a_i$ from state $s_i$, sometimes also written with a time index $t$.
The {\em intrinsic reward} $u(\mathbf{f})$ is an aggregate social preference across features $\mathbf{f}$ and is calculated according to the formula,
\begin{align}\label{eq:inequityextend}
u_{i}(\mathbf{f}_i | \boldsymbol{\theta}) = \mathbf{v}^\mathrm{T} \sigma \left ( \mathbf{W}^\mathrm{T} \mathbf{f}_i  + \mathbf{b} \right )  \,,
\end{align}
where $\sigma$ is the ReLU activation function, and $\boldsymbol{\theta} = \{\mathbf{W}, \mathbf{v}, \mathbf{b}\}$ are the parameters of a 2-layer neural network with 2 hidden nodes. These parameters are evolved based on fitness (see Section \ref{sec:arch_training}). The elements of $\mathbf{v} = (v_1, v_2)$ approximately correspond to a linear combination of the coefficients related to advantagenous and disadvantagenous inequity aversion mentioned in \cite{hughes2018inequity}, which were found via grid search in this previous work, but are here evolved.

\begin{figure*}[htb]
    \centering
    \includegraphics[width=1.\textwidth]{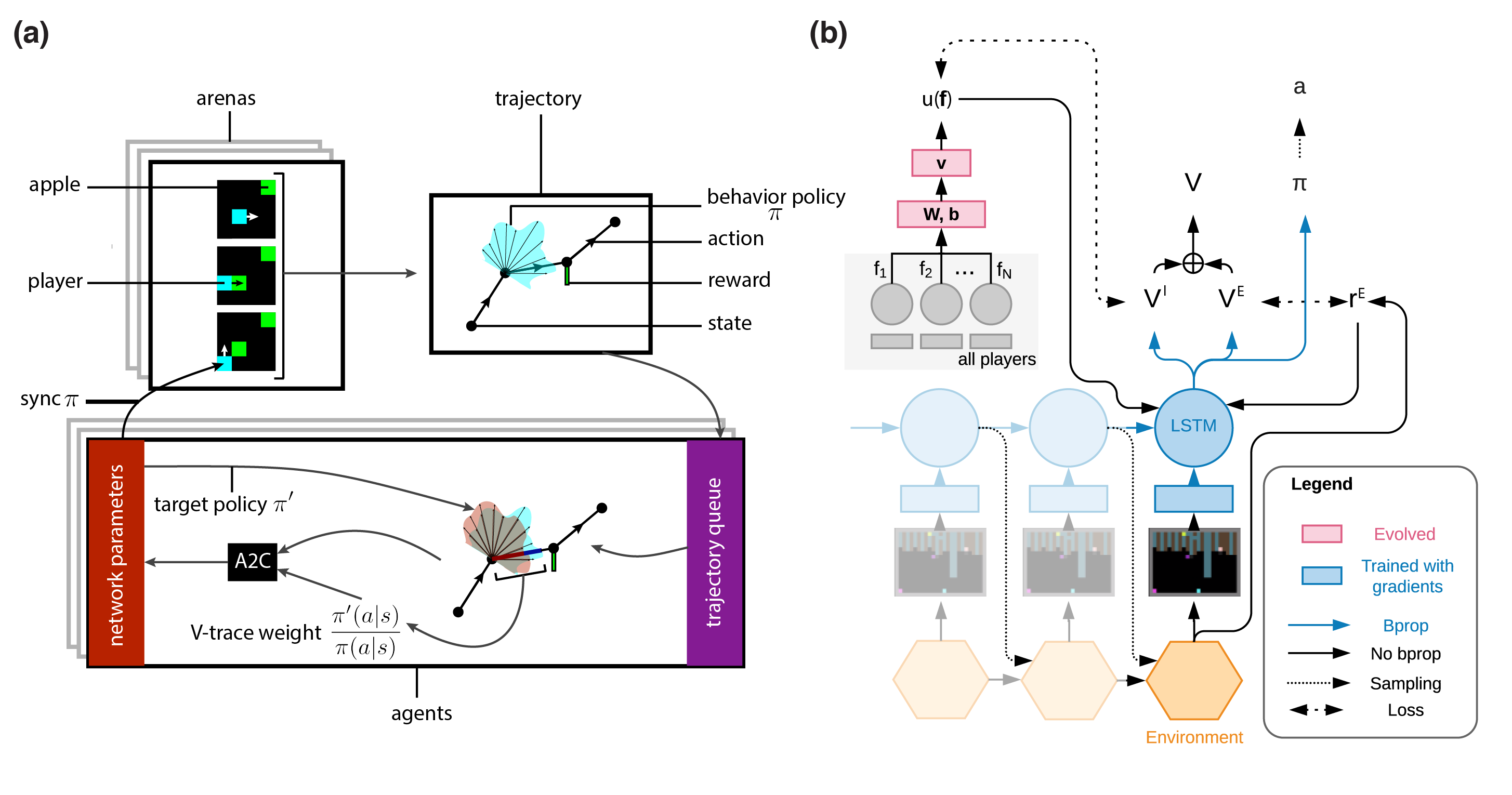}
    \vspace{-1.0cm}
\caption{(a) Agent $A_j$ adjusts policy $\pi_j(s,a|\phi)$ using off-policy importance weighted actor-critic (V-Trace) \citep{espeholt2018impala} by sampling from a queue with (possibly stale) trajectories recorded from 500 actors acting in parallel arenas. (b) The architecture (shown only for 1 agent) includes a visual encoder (1-layer convolutional neural net with 6 3x3 filters, stride 1, followed by two fully-connected layers with 32 units each), intrinsic and extrinsic value heads ($V^I$ and $V^E$), a policy head $\pi$, and a long-short term memory (LSTM, with 128 hidden units), which takes last intrinsic and extrinsic rewards ($u(\mathbf{f})$ and $r^E$) and last action as input. The reward network weights are evolved based on total episode return.}
\label{fig:arch}
\vspace{-.25cm}
\end{figure*}


\newcommand{\task}{\mathcal{T}}
\newcommand{\data}{\mathcal{D}}
\newcommand{\loss}{\mathcal{L}}
\newcommand{\fitness}{\mathcal{F}}
\newcommand{\population}{\mathcal{P}}
\newcommand{\popupdate}{\mathcal{U}}
\newcommand{\getpopbatch}{\mathcal{S}}
\newcommand{\lossi}{\loss_{\task_i}}
\newcommand{\params}{\theta}
\newcommand{\hyperparams}{h}

\makeatletter
\def\BState{\State\hskip-\ALG@thistlm}
\makeatother


\begin{algorithm}[t]
\caption{Training Pseudocode - Shared Reward Network}
\label{alg:pseudocode}
\begin{algorithmic}[1]
\small
\Require $\mathcal{P}$: pop. of policy networks and hyper-parameters $\{\{\phi_1,h_1\}, ..., \{\phi_N, h_N\}\}$
\Require $\mathcal{R}$: pop. of reward networks $\{\theta_1, ..., \theta_N\}$, where  $\theta=\{\mathbf{W},\mathbf{v},\mathbf{b}\}$
\Require $\getpopbatch$: procedure to sample from $\mathcal{P}$ and $\mathcal{R}$ and return 5 players
\Require $\popupdate$: procedure to update/evolve weights given a population of fitness-scored individuals
\Require $\fitness$: procedure to calculate and assign fitness given the sampled players
\While{not done}
\State $\mathrm{p_{1:5}} = \mathrm{p_1},...,\mathrm{p_5} \leftarrow \getpopbatch(\mathcal{P}, \mathcal{R})$ \Comment{Sample 5 players}
\State $\mathbf{env} \sim p(\task)$ \Comment{Sample from distribution $p$ over environments $\task$}
\State $e_{1:5}^0 \leftarrow 0$  \Comment{Initialize temporally decayed reward}
\For {$t:=1$ \textbf{to} $T$} \Comment{Run for $T=1000$ steps}
\State $\tau^t = (r_{1:5}^{E,t}, V_{1:5}^{E,t}, o_{1:5}^t, a_{1:5}^t) \leftarrow \mathbf{env}(p_{1:5})$ \Comment{Play players in environment, collect outputs in a trajectory $\tau=\{\tau^1,...,\tau^T\}$}
\For {$j:=1$ \textbf{to} $5$} \Comment {Calculate feature vectors for players}
\If {\textbf{retrospective}} 
\State $f_{j}=e_j^t \leftarrow \eta \,e_j^{t-1} + r_j^{E,t} $ \Comment{Decayed extrinsic reward}
\ElsIf {\textbf{prospective}} 
\State $f_{j} \leftarrow V_j^{E,t}$   \Comment{Value estimate from extrinsic value head}
\EndIf
\EndFor
\For {$i:=1$ \textbf{to} $5$} \Comment {Calculate intrinsic rewards for players}
\State $\mathbf{f}_i \leftarrow \mathbf{reorder}(\mathbf{f_i},i)$ 
\State $u^t_i \leftarrow \mathbf{v}^\mathrm{T} \sigma \left ( \mathbf{W}^\mathrm{T} \mathbf{f}_i  + \mathbf{b} \right )$ \Comment{Calculate intrinsic reward}
\EndFor
\EndFor
\For {$i:=1$ \textbf{to} $5$} 
\State $\phi_i \leftarrow \mathbf{RL}(\phi_i, \tau, u_i, h_i)$  \Comment{RL training for each $\phi$}
\State $F_{\phi_k}, F_{\theta_k} \leftarrow \fitness (\tau)$  \Comment{Calculate smoothed fitnesses associated with each reward and policy network sampled in this episode}
\EndFor
\For {$(\phi_k, h_k) \in \mathcal{P}$}
\If {\textbf{available\_to\_evolve}$(\phi_k, h_k)$} \Comment{If burn-in period has passed, update population based on smoothed fitness of individuals}
\State $(\phi_k, h_k) \leftarrow \popupdate(\mathcal{P}, F_{\phi_k}$)
\EndIf
\EndFor
\For {$\theta_k \in \mathcal{R}$}
\If {\textbf{available\_to\_evolve}$(\theta_k)$} 
\State $\theta_k \leftarrow \popupdate(\mathcal{R}, F_{\theta_k}$)
\EndIf
\EndFor
\EndWhile
\end{algorithmic}
\end{algorithm}


The feature vector $\mathbf{f}_i$ is a player-specific vector quantity that agents can transform into intrinsic reward via their reward network. It's composed of features $f_{ij}$ derived from all players \footnote{Note that we use both $i$ and $j$ to index over the players, but $i$ makes reference to the player \emph{receiving} the intrinsic reward, while $j$ indexes the players \emph{sending} the features over which the intrinsic reward of player $i$ is defined.}, so that each player has access to the same set of features, with the exception that its own feature is demarcated specially (by always occupying the first element of the vector). The features themselves are a function of recently received or expected future (extrinsic) reward for each agent. In Markov games the rewards received by different players may not be aligned in time. Thus, any model of social preferences should not be overly influenced by the precise temporal alignment of different players' rewards.  Intuitively, they ought to depend on comparing temporally averaged reward estimates between players, rather than instantaneous values. Therefore, we considered two different ways of temporally aggregating the rewards.

\begin{figure*}[htb]
    \centering
    \includegraphics[width=0.85\textwidth]{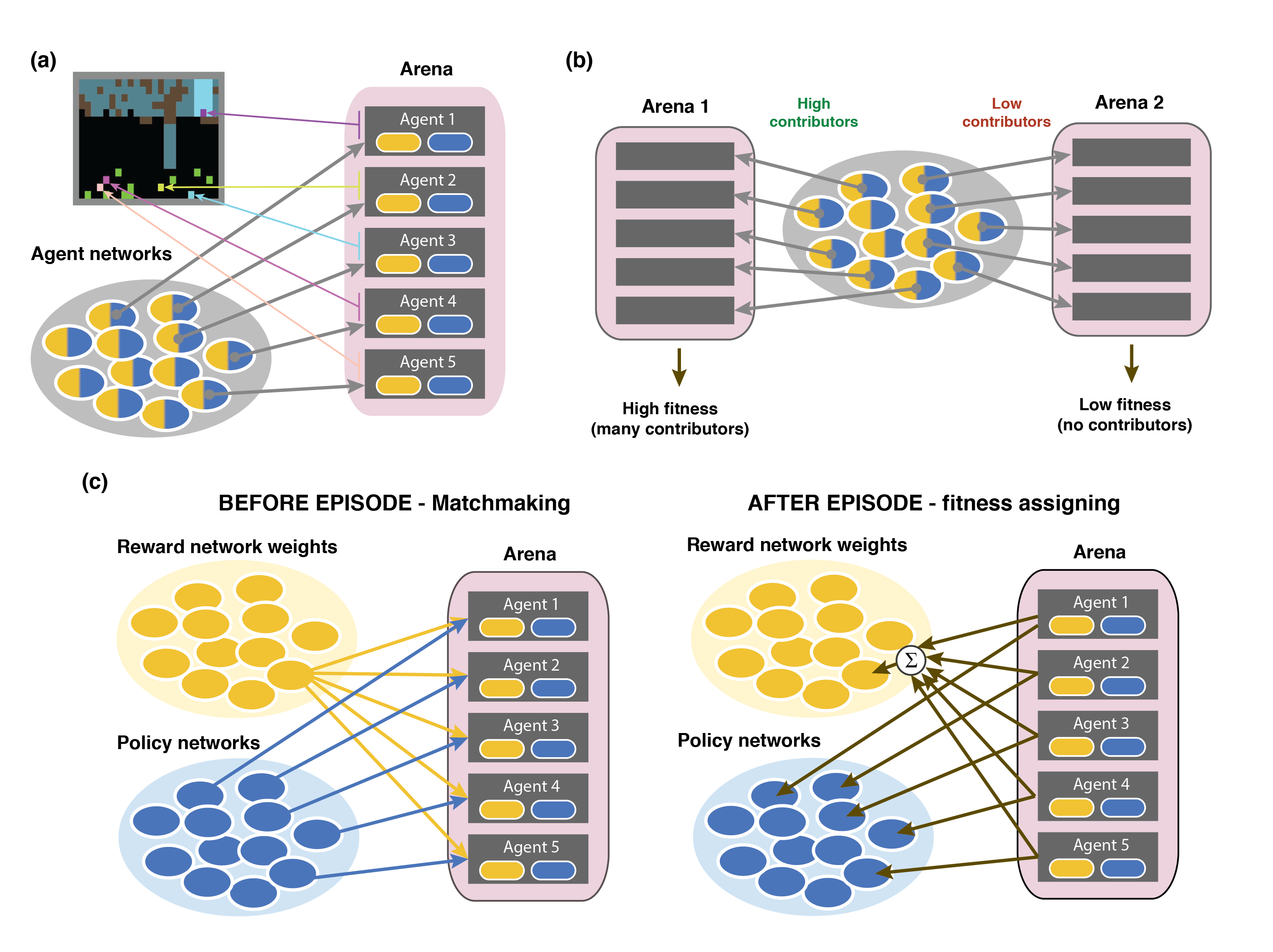}
    \vspace{-.5cm}
\caption{(a) Agents assigned and evolved with individual reward networks. (b) Assortative matchmaking, which preferentially plays cooperators with other cooperators and defectors with other defectors. (c) A single reward network is sampled from the population and assigned to all players, while 5 policy networks are sampled and assigned to the 5 players individually. After the episode, policy networks evolve according to individual player returns, while reward networks evolve according to aggregate returns over all players.}
\label{fig:evo}
\vspace{-.25cm}
\end{figure*}

The {\em retrospective} method derives intrinsic reward from whether an agent judges that other agents have been actually (extrinsically) rewarded in the recent past. The {\em prospective} variant derives intrinsic reward from whether other agents are expecting to be (extrinsically) rewarded in the near future.\footnote{Our terms prospective and retrospective map onto the terms intentional and consequentialist respectively as used by \cite{lerer17, peysakhovich2018}.} For the retrospective variant, $f_{ij} = e^t_j$, where the temporally decayed reward $e_j^t$ for the agents $j = 1,\dots, N$ are updated at each timestep $t$ according to
\begin{equation}
     e_j^t= \eta \,e_j^{t-1} + r_j^{E,t} \, ,
\end{equation}
and $\eta = 0.975$. The prospective variant uses the value estimates $V^E_j$ (see Figure \ref{fig:arch}b) for $f_{ij}$ and has a stop-gradient before the reward network module so that gradients don't flow back into other agents (as in for example DIAL from \citep{foerster2016}).

\subsection{Architecture and Training}
\label{sec:arch_training}

We used the same training framework as in \cite{jaderberg2018human}, which performs distributed asynchronous training in multi-agent environments, including population-based training (PBT) \citep{jaderberg2017population}. We trained a population of $N=50$ agents\footnote{Similar to as in \citep{espeholt2018impala}, we distinguish between an "agent" which acts in the environment according to some policy, and a "learner" which updates the parameters of a policy. In principle, a single agent's policy may depend on parameters updated by several separate learners.} with policies $\{\pi_i\}$, from which we sampled $5$ players in order to populate each of $500$ arenas (where \emph{arena} is an instantiation of a single episode of the environment) running in parallel. Within each arena, an episode of the environment was played with the sampled agents, before resampling new ones. Agents were sampled using one of two matchmaking processes (described in more detail below). Episode trajectories lasted 1000 steps and were written to queues for learning, from which weights were updated using V-Trace (Figure \ref{fig:arch}a).

The set of weights evolved included learning rate, entropy cost weight, and reward network weights $\theta$\footnote{We can imagine that the reward weights are simply another set of optimization hyperparameters since they enter into the loss.}. The parameters of the policy network $\phi$ were inherited in a Lamarckian fashion as in \citep{jaderberg2017population}.  Furthermore, we allowed agents to observe their last actions $a_{i,t-1}$, last intrinsic rewards ($u_{i,t-1}(\mathbf{f}_i)$), and last extrinsic rewards ($r_{i,t-1}^E(s_i,a_i)$) as input to the LSTM in the agent's neural network. 

The objective function was identical to that presented in \cite{espeholt2018impala} and comprised three components: (1) the value function gradient, (2) policy gradient, and (3) entropy regularization, weighted according to hyperparameters baseline cost and entropy cost (see Figure \ref{fig:arch}b). 

Evolution was based on a fitness measure calculated as a moving average of total episode return, which was a sum of apples collected minus penalties due to tagging, smoothed as follows:
\begin{equation}
     F_i^{n}= (1-\nu) F_i^{n-1} + \nu R_i^{n} \, ,
\end{equation}
where $\nu = 0.001$ and $R_i^n = \sum_t r^{E,t}_{i}$ is the total episode return obtained on episode $n$ by agent $i$ (or reward network $i$ in the case of the shared reward network evolution (see Section \ref{sec:multi_evolve} for details).

Training was done via joint optimization of network parameters via SGD and hyperparameters/reward network parameters via evolution in the standard PBT setup. Gradient updates were applied for every trajectory up to a maximum length of 100 steps, using a batch size of 32. Optimization was via RMSProp with epsilon=$10{^{-5}}$, momentum=0, decay rate=0.99, and an RL discount factor of 0.99. The baseline cost weight (see \citet{mnih2016}) was fixed at 0.25, and the entropy cost was sampled from LogUniform($2\times10{^{-4}}$,0.01) and evolved throughout training using PBT. The learning rates were all initially set to $4\times10{^{-4}}$ and then allowed to evolve.

PBT uses evolution (specifically genetic algorithms) to search over a space of hyperparameters rather than manually tuning or performing a random search, resulting in an adaptive schedule of hyperparameters and joint optimization with network parameters learned through gradient descent \cite{jaderberg2017population}. 

There was a mutation rate of $0.1$ when evolving hyperparameters, using multiplicative perturbations of $\pm 20\%$ for entropy cost and learning rate, and additive perturbation of $\pm 0.1$ for reward network parameters.  We implemented a burn-in period for evolution of $4\times10^6$ agent steps, to allow network parameters and hyperparameters to be used in enough episodes for an accurate assessment of fitness before evolution.

\subsection{Random vs. assortative matchmaking}

Matches were determined according to two methods: (1) random matchmaking and (2) assortative matchmaking. Random matchmaking simply selected uniformly at random from the pool of agents to populate the game, while cooperative matchmaking first ranked agents within the pool according to a metric of recent cooperativeness, and then grouped agents such that players of similar rank played with each other. This ensured that highly cooperative agents played only with other cooperative agents, while defecting agents played only with other defectors. For Cleanup, cooperativeness was calculated based on the amount of steps in the last episode the agent chose to clean. For Harvest, it was calculated based on the difference between the the agent's return and the mean return of all players, so that having less return than average yielded a high cooperativeness ranking. Cooperative metric-based matchmaking was only done with either individual reward networks or no reward networks (Figure \ref{fig:evo}b). We did not use cooperative metric-based matchmaking for our multi-level selection model, since these are theoretically separate approaches.

\begin{figure*}[htb]
    \centering
    \includegraphics[width=0.9\textwidth]{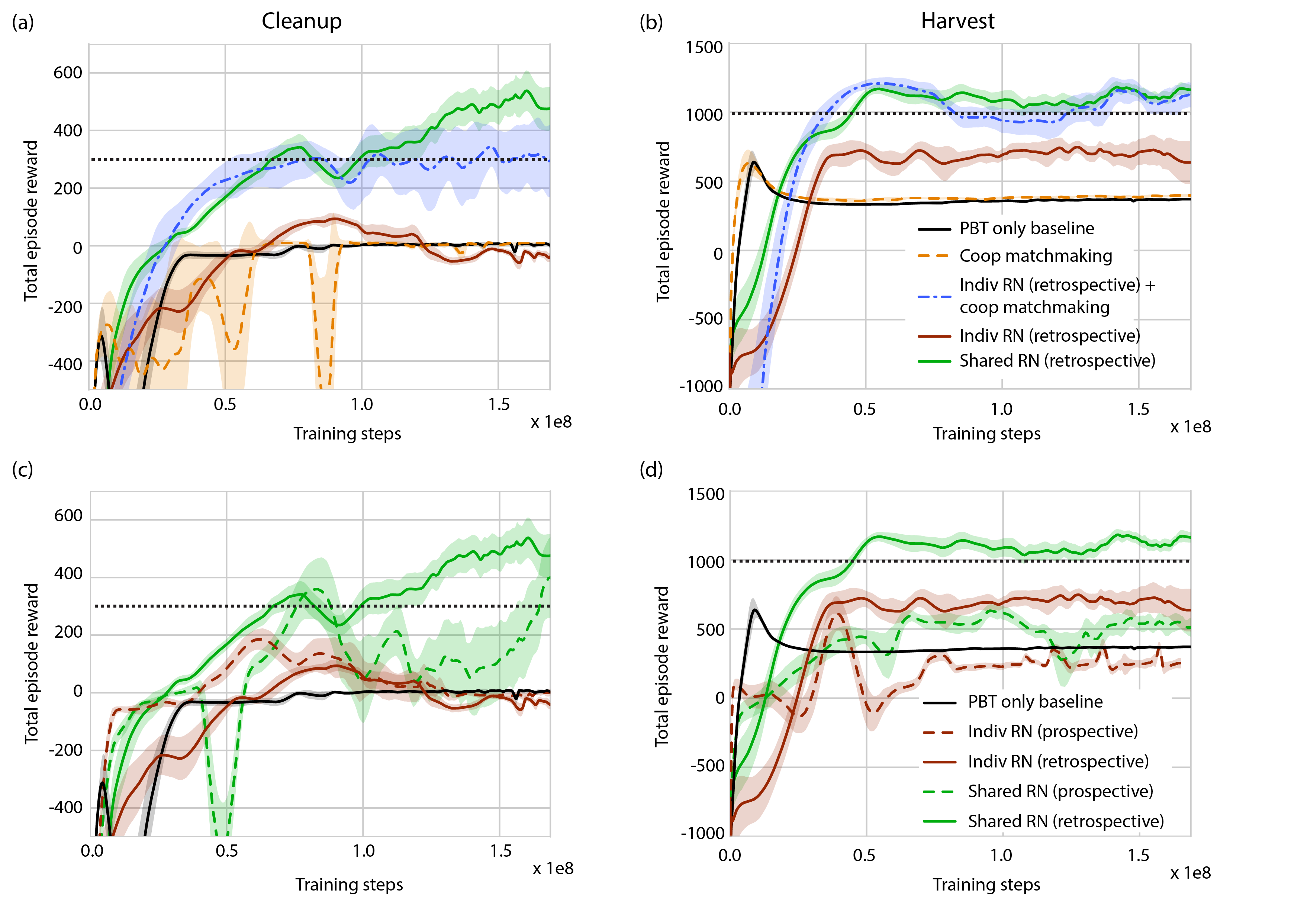}
    \vspace{-.5cm}
\caption{Total episode rewards, aggregated over players. (a), (b): Comparing retrospective (backward-looking) reward evolution with assortative matchmaking and PBT-only baseline in (a) Cleanup and (b) Harvest. (c), (d): Comparing prospective (forward-looking) with retrospective (backward-looking) reward evolution in (c) Cleanup and (d) Harvest. The black dotted line indicates performance from \cite{hughes2018inequity}. The shaded region shows standard error of the mean, taken over the population of agents.}
\label{fig:results1}
\vspace{-.25cm}
\end{figure*}

\subsection{Individual vs. shared reward networks}
\label{sec:multi_evolve}

Building on previous work that evolved either the intrinsic reward \citep{jaderberg2018human} or the entire loss function \citep{houthooft2018evolved}, we considered the reward network weights to be hyperparameters that could be evolved in parallel with the policy parameters (Figure \ref{fig:evo}a). Distinct from these methods, we separately evolved the reward network within its own population, thereby allowing different modules of the agent to compete only with like components. This allowed for independent exploration of hyperparameters via separate credit assignment of fitness, and thus considerably more of the hyperparameter landscape could be explored compared with using only a single pool. In addition, reward networks could be randomly assigned to any policy network, and so were forced to generalize to a wide range of policies. In a given episode, $5$ separate policy networks were paired with the same reward network, which we term a {\em shared reward network}. In line with \citep{jaderberg2017population}, the fitness determining the copying of policy network weights and evolution of optimization-related hyperparameters (entropy cost and learning rate) were based on individual agent return. By contrast, the reward network parameters were evolved according to fitness based on total episode return across the group of co-players (Figure \ref{fig:evo}c).

This contribution is distinct from previous work which evolved intrinsic rewards \citep[e.g.][]{jaderberg2018human} because (1) we evolve over social features rather than a remapping of environmental events, and (2) reward network evolution is motivated by dealing with the inherent tension in ISDs, rather than merely providing a denser reward signal. In this sense it's more akin to evolving a form of communication for social cooperation, rather than learning reward-shaping in a sparse-reward environment. We allow for multiple agents to share the same components, and as we shall see, in a social setting, this winds up being critical. Shared reward networks provide a biologically principled method that mixes group fitness on a long timescale and individual reward on a short timescale. This contrasts with hand-crafted means of aggregation, as in previous work \citep{chang2004all,mataric1994learning}.

\section{Results}

As shown in Figure \ref{fig:results1}, PBT without using an intrinsic reward network performs poorly on both games, where it asymptotes to 0 total episode reward in Cleanup and 400 for Harvest (the number of apples gained if all agents collect as quickly as they can). 

Figures \ref{fig:results1}a-b compare random and assortative matchmaking with PBT and reward networks using retrospective social features. When using random matchmaking, individual reward network agents perform no better than PBT at Cleanup, and only moderately better at Harvest.  Hence there is little benefit to adding reward networks over social features if players have separate networks, as these tend to be evolved selfishly. The assortative matchmaking experiments used either no reward network ($u(\mathbf{f})$ = 0) or individual reward networks. Without a reward network, performance was the same as the PBT baseline. With individual reward networks, performance was very high, indicating that both conditioning the internal rewards on social features and a preference for cooperative agents to play together were key to resolving the dilemma. On the other hand, shared reward network agents perform as well as assortative matchmaking and the handcrafted inequity aversion intrinsic reward from \citep{hughes2018inequity}, even using random matchmaking. This implies that agents didn't necessarily need to have immediate access to honest signals of other agents' cooperativeness to resolve the dilemma; it was enough to simply have the same intrinsic reward function, evolved according to collective episode return. Videos comparing performance of the PBT baseline with the retrospective variant of shared reward network evolution can be found at \href{https://www.youtube.com/watch?v=medBBLLM4c0}{https://youtu.be/medBBLLM4c0} and \href{https://www.youtube.com/watch?v=yTjrlH3Ms9U}{https://youtu.be/yTjrlH3Ms9U}.

Figures \ref{fig:results1}(c) and (d) compare the retrospective and prospective variants of reward network evolution. The prospective variant, although better than PBT when using a shared reward network, generally results in worse performance and more instability. This is likely because the prospective variant depends on agents learning good value estimates before the reward networks become useful, whereas the retrospective variant only depends on environmentally provided reward and thus does not suffer from this issue. Interestingly, we observed that the prospective variant does achieve very high performance if gradients are allowed to pass between agents via the value estimates $V^E_j$ (data not shown); however, this constitutes centralized learning, albeit with decentralized execution (see \cite{foerster2016}). Such approaches are promising but less consistent with biologically plausible mechanisms of multi-agent learning which are of interest here and so were not pursued.

\begin{figure*}[htb]
    \centering
    \includegraphics[width=0.9\textwidth]{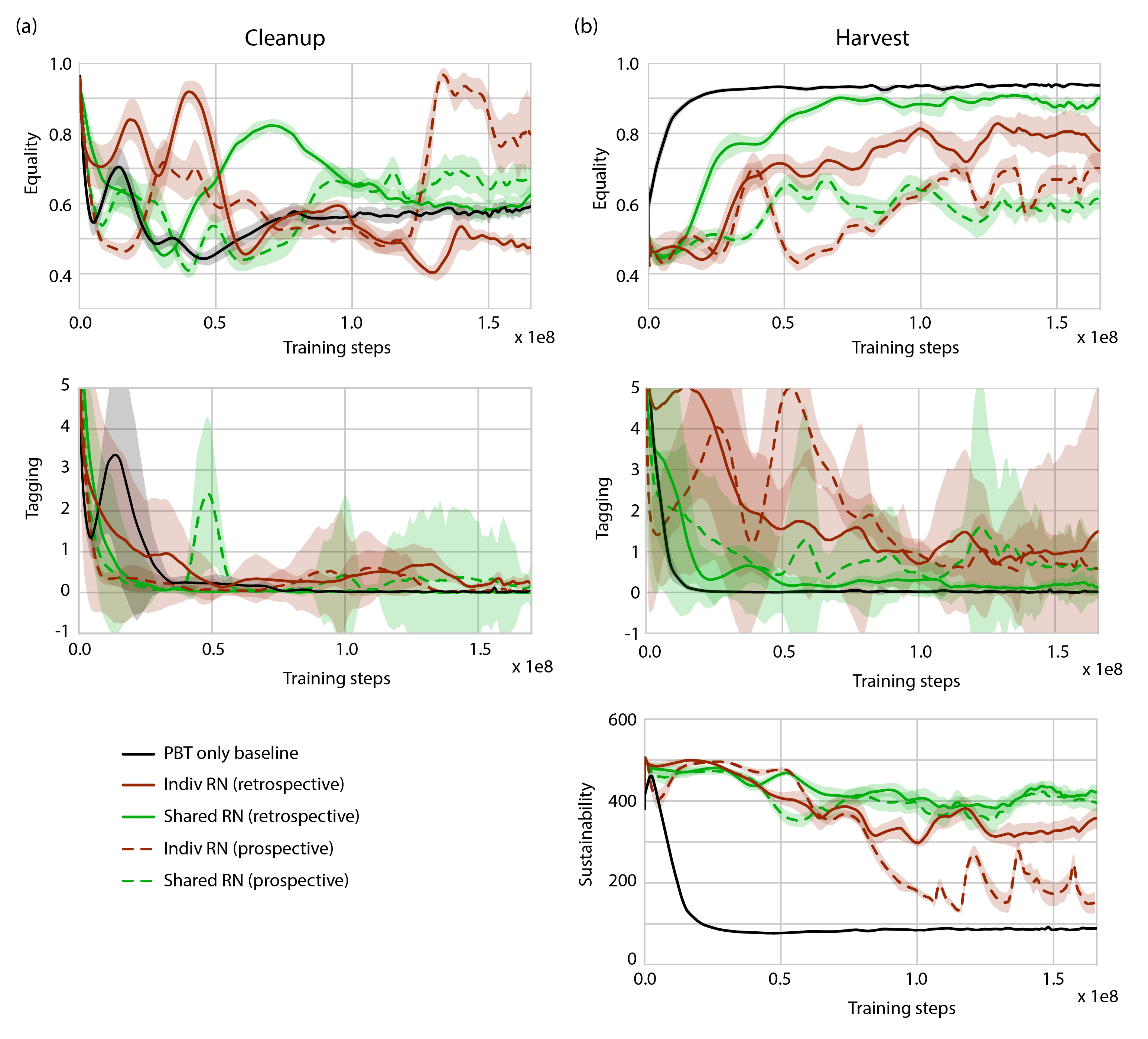}
    \vspace{-.25cm}
\caption{Social outcome metrics for (a) Cleanup and (b) Harvest. \textit{Top:} equality, \textit{middle:} total amount of tagging, \textit{bottom:} sustainability. The shaded region shows the standard error of the mean.}
\label{fig:results2}
\vspace{-.25cm}
\end{figure*}

We next plot various social outcome metrics in order to better capture the complexities of agent behavior (see Figure \ref{fig:results2}). 
Equality is calculated as $\mathbb{E}(1-G(\mathbf{R}))$, where $G(\mathbf{R})$ is the Gini coefficient over individual returns. Figure \ref{fig:results2}b demonstrates that, in Harvest, having the prospective version of reward networks tends to lead to lower equality, while the retrospective variant has very high equality. Equality in Cleanup is more unstable throughout training, since it's not necessarily optimal, but tends to be lower overall than for Harvest, even when performance is high, indicating that equality might be harder to achieve in different games.
Tagging measures the average number of times a player fined another player throughout the episode. The middle panel of Figure \ref{fig:results2}b shows that there is a higher propensity for tagging in Harvest when using either a prospective reward network or an individual reward network, compared to the retrospective shared reward network. This explains the performance shown in Figure \ref{fig:results1}, as being tagged results in a very high negative reward. Tagging in the Cleanup task is overall much lower than in Harvest.
Sustainability measures the average time step on which agents received positive reward, averaged over the episode and over agents. We see in the bottom panel of \ref{fig:results2}b that having no reward network results in players collecting apples extremely quickly in Harvest, compared with much more sustainable behavior with reward networks. In Cleanup, the sustainability metric is not meaningful and so this was not plotted.

Finally, we can directly examine the weights of the final retrospective shared reward networks which were best at resolving the ISDs. Interestingly, the final weights evolved in the second layer suggest that resolving each game might require a different set of social preferences. In Cleanup, one of the final layer weights $v_2$ evolved to be close to $0$, whereas in Harvest, $v_1$ and $v_2$ evolved to be of large magnitude but opposite sign. We can see a similar pattern with the biases $\mathbf{b}$. We interpret this to mean that Cleanup required a less complex reward network: it was enough to simply find other agents' being rewarded as intrinsically rewarding. In Harvest, however, a more complex reward function was perhaps needed in order to ensure that other agents were not over-exploiting the apples. We found that the first layer weights $\mathbf{W}$ tended to take on arbitrary (but positive) values. This is because of random matchmaking: co-players were randomly selected and thus there was little evolutionary pressure to specialize these weights.

\begin{figure}[htb]
    \centering
    \includegraphics[width=0.48\textwidth]{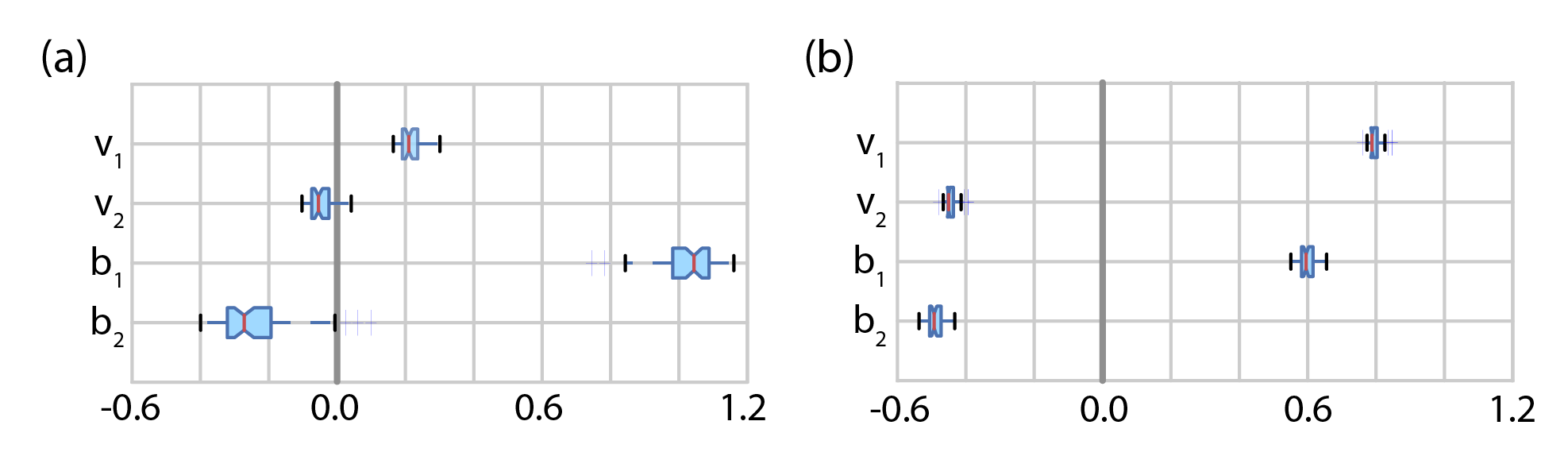}
\caption{Distribution of layer 2 weights and biases of evolved retrospective shared reward network at $1.5\times10{^{8}}$ training steps for (a) Cleanup, and (b) Harvest.}
\vspace{-0.3cm}
\label{fig:results3}
\end{figure}

\section{Discussion}

Real environments don't provide scalar reward signals to learn from. Instead, organisms have developed various internal drives based on either primary or secondary goals \citep{Baldassarre13}. Here we examined intrinsic rewards based on features derived from other agents in the environment, in order to establish whether such social signals could enable the evolution of altruism to solve intertemporal social dilemmas. In accord with evolutionary theory \citep{axelrod81, Nowak1560}, we found that na\"ively implementing natural selection via genetic algorithms did not lead to the emergence of cooperation. Furthermore, assortative matchmaking was sufficient to generate cooperative behavior in cases where honest signals were available. Finally, we proposed a new multi-level evolutionary paradigm based on shared reward networks that achieves cooperation in more general situations.

We demonstrated that the reward network weights evolved differently for Cleanup versus Harvest, indicating that the two tasks necessitate different forms of social cooperation for optimal performance. This highlights the advantage of evolving rather than hand-crafting the weighting between individual reward and group reward, as optimal weightings cannot necessarily be anticipated for all environments. Evolving such weightings thus constitutes a form of meta-learning, wherein an entire learning system, including intrinsic reward functions, is optimized for fast learning \cite{singh2010intrinsically, fernando2018meta}. Here we have extended these ideas to the multi-agent domain.

Why does evolving intrinsic social preferences promote cooperation? Firstly, evolution ameliorates the intertemporal choice problem by distilling the long timescale of collective fitness into the short timescale of individual reinforcement learning, thereby improving credit assignment between selfish acts and their temporally displaced negative group outcomes \citep{hughes2018inequity}. Secondly, it mitigates the social dilemma itself by allowing evolution to expose social signals that correlate with, for example, an agent's current level of selfishness. Such information powers a range of mechanisms for achieving mutual cooperation like competitive altruism \citep{hardy2006nice}, other-regarding preferences \citep{cooper2016other}, and inequity aversion \citep{fehr1999}. In accord, laboratory experiments show that humans cooperate more readily when they can communicate \citep{ostrom1992covenants, janssen2010lab}. 

The shared reward network evolution model was inspired by multi-level selection; yet it does not correspond to the prototypical case of that theory since its lower level units of evolution (the policy networks) are constantly swapping which higher level unit (reward network) they are paired with. Nevertheless, there are a variety of ways in which we see this form of modularity arise in nature. For example, free-living  microorganisms occasionally form multi-cellular structures to solve a higher order adaptive problem, like slime mold forming a spore-producing stalk for dispersal \citep{west2006social}, and many prokaryotes can incorporate plasmids (modules) found in their environment or received from other individuals as functional parts of their genome, thereby achieving cooperation in social dilemmas  \citep{griffin2004cooperation, mc2011horizontal}. Alternatively, in humans a reward network may represent a shared ``cultural norm'', with its fitness based on cultural information accumulated from the groups in which it holds sway. In this way, the spread of norms can occur independently of the success of individual agents \citep{boyd2009}.

Note that in this work, we have assumed that agents have perfect knowledge of other agents' rewards, while in real-world systems this is not typically the case. This assumption was made in order to disentangle the effects of cultural evolution from the quality of the signals being evolved over. Natural next steps include adding partial observability or noise to this signal (to make it more analogous to, for instance, a smile/frown or other locally observable social signals), identifiability across episodes, or even deception.

The approach outlined here opens avenues for investigating alternative evolutionary mechanisms for the emergence of cooperation, such as kin selection \citep{griffin2002} and reciprocity \citep{trivers1971evolution}. It would be interesting to see whether these lead to different weights in a reward network, potentially hinting at the evolutionary origins of different social biases. Along these lines, one might consider studying an emergent version of the assortative matchmaking model along the lines suggested by \cite{henrich2003}, adding further generality and power to our setup. Finally, it would be fascinating to determine how an evolutionary approach can be combined with multi-agent communication to produce that most paradoxical of cooperative behaviors: cheap talk.

\begin{acks}
We would like to thank Simon Osindero, Iain Dunning, Andrea Tacchetti, and many DeepMind colleagues for valuable discussions and feedback, as well as code development and support. 
\end{acks}

\bibliographystyle{ACM-Reference-Format}  

\balance  

\bibliography{references}

\end{document}